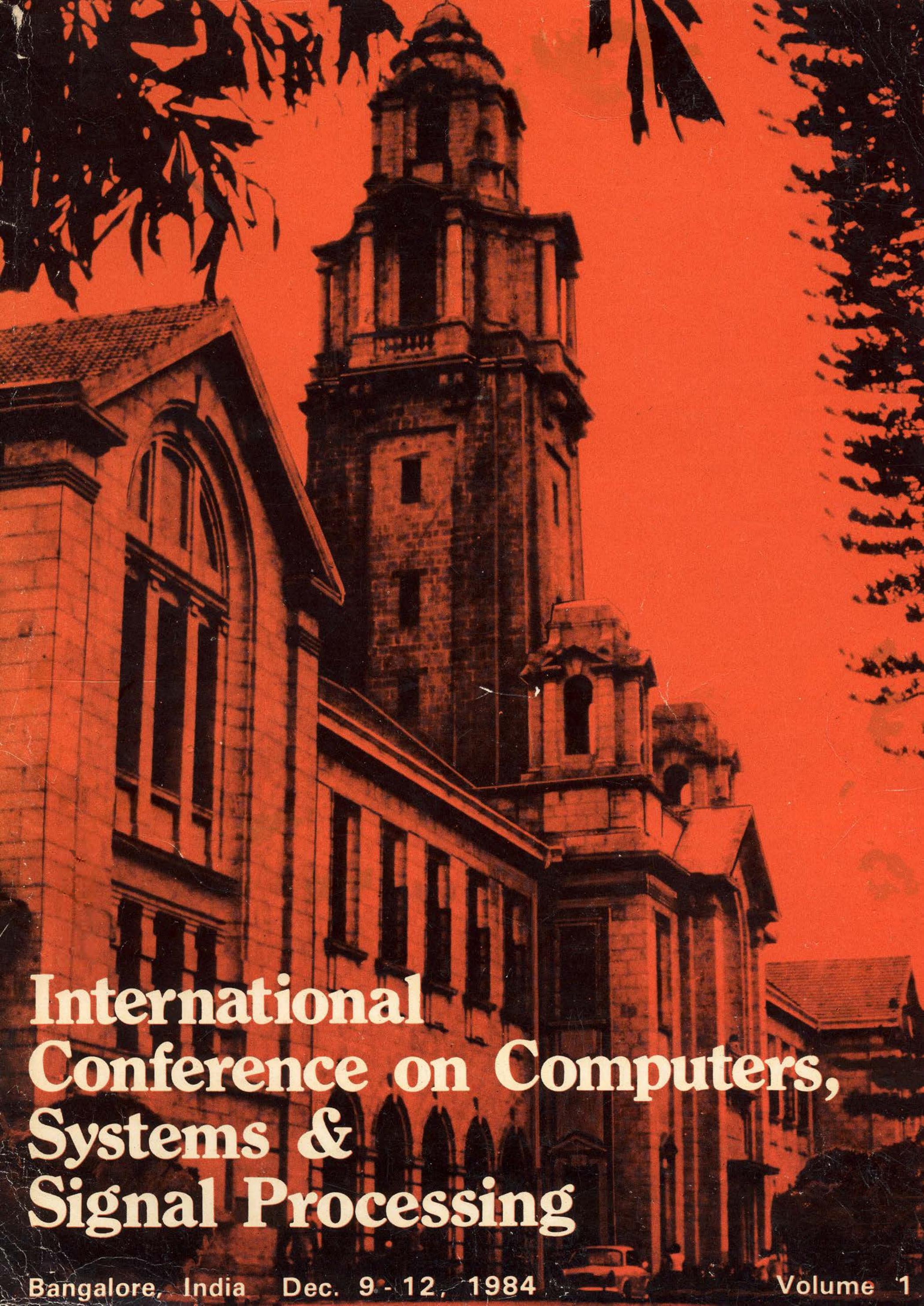

# International Conference on Computers, Systems & Signal Processing

Bangalore, India    Dec. 9 - 12, 1984    Volume 1

# QUANTUM CRYPTOGRAPHY: PUBLIC KEY DISTRIBUTION AND COIN TOSSING

Charles H. Bennett (IBM Research, Yorktown Heights NY 10598 USA)
Gilles Brassard (dept. IRO, Univ. de Montreal, H3C 3J7 Canada)

When elementary quantum systems, such as polarized photons, are used to transmit digital information, the uncertainty principle gives rise to novel cryptographic phenomena unachieveable with traditional transmission media, e.g. a communications channel on which it is impossible in principle to eavesdrop without a high probability of disturbing the transmission in such a way as to be detected. Such a quantum channel can be used in conjunction with ordinary insecure classical channels to distribute random key information between two users with the assurance that it remains unknown to anyone else, even when the users share no secret information initially. We also present a protocol for coin-tossing by exchange of quantum messages, which is secure against traditional kinds of cheating, even by an opponent with unlimited computing power, but, ironically can be subverted by use of a still subtler quantum phenomemon, the Einstein-Podolsky-Rosen paradox.

I. Introduction

Conventional cryptosystems such as ENIGMA, DES, or even RSA, are based on a mixture of guesswork and mathematics. Information theory shows that traditional secret-key cryptosystems cannot be totally secure unless the key, used once only, is at least as long as the cleartext. On the other hand, the theory of computational complexity is not yet well enough understood to prove the computational security of public-key cryptosystems.

In this paper we use a radically different foundation for cryptography, viz. the uncertainty principle of quantum physics. In conventional information theory and cryptography, it is taken for granted that digital communications in principle can always be passively monitored or copied, even by someone ignorant of their meaning. However, when information is encoded in non-orthogonal quantum states, such as single photons with polarization directions 0, 45, 90, and 135 degrees, one obtains a communications channel whose transmissions in principle cannot be read or copied reliably by an eavesdropper ignorant of certain key information used in forming the transmission. The eavesdropper cannot even gain partial information about such a transmission without altering it a random and uncontrollable way likely to be detected by the channel's legitimate users.

Quantum coding was first described in [W], along with two applications: making money that is in principle impossible to counterfeit, and multiplexing two or three messages in such a way that reading one destroys the others. More recently [BBBW], quantum coding has been used in conjunction with public key cryptographic techniques to yield several schemes for unforgeable subway tokens. Here we show that quantum coding by itself achieves one of the main advantages of public key cryptography by permitting secure distribution of random key information between parties who share no secret information initially, provided the parties have access, besides the quantum channel, to an ordinary channel susceptible to passive but not active eavesdropping. Even in the presence of active eavesdropping, the two parties can still distribute key securely if they share some secret information initially, provided the eavesdropping is not so active as to suppress communications completely. We also present a protocol for coin tossing by exchange of quantum messages. Except where otherwise noted the protocols are provably secure even against an opponent with superior technology and unlimited computing power, barring fundamental violations of accepted physical laws.

Offsetting these advantages is the practical disadvantage that quantum transmissions are necessarily very weak and cannot be amplified in transit. Moreover, quantum cryptography does not provide digital signatures, or applications such as certified mail or the ability to settle disputes before a judge.

II. Essential Properties of Polarized Photons

Polarized light can be produced by sending an ordinary light beam through a polarizing apparatus such as a Polaroid filter or calcite crystal; the beam's polarization axis is determined by the orientation of the polarizing apparatus in which the beam originates. Generating single polarized photons is also possible, in principle by picking them out of a polarized beam, and in practice by a variation of an experiment [AGR] of Aspect, et. al.

Although polarization is a continuous variable, the uncertainty principle forbids measurements on any single photon from revealing more than one bit about its polarization. For example, if a light beam with polarization axis $\alpha$ is sent into a filter oriented at angle $\beta$, the individual photons behave dichotomously and probabilistically, being transmitted with probability $\cos^2(\alpha-\beta)$ and absorbed with the complementary probability $\sin^2(\alpha-\beta)$. The photons behave deterministically only when the two axes are parallel (certain transmission) or perpendicular (certain absorbtion).

If the two axes are not perpendicular, so that some photons are transmitted, one might hope to learn additional information about $\alpha$ by measuring the transmitted photons again with a polarizer oriented at some third angle; but this is to no avail, because the transmitted photons, in passing through the $\beta$ polarizer, emerge with exactly $\beta$ polarization, having lost all memory of their previous polarization $\alpha$.

Another way one might hope to learn more than one bit from a single photon would be not to measure it directly, but rather somehow amplify it into a clone of identically polarized photons, then perform measurements on these; but this hope is also vain, because such cloning can be shown to be inconsistent with the foundations of quantum mechanics [WZ].

Formally, quantum mechanics represents the internal state of a quantum system (e.g. the polarization of a photon) as a vector $\psi$ of unit length in a linear space H over the field of complex numbers (Hilbert space). The inner product of two vectors $<\phi|\psi>$, is defined as $\Sigma_j \phi_j^* \psi_j$, where * indicates complex conjugation. The dimensionality of the Hilbert space depends on the system, being larger (or even infinite) for more complicated systems. Each physical measurement M that might be performed on the system corresponds to a resolution of its Hilbert space into orthogonal subspaces, one for each possible outcome of the measurement. The number of possible outcomes is thus limited to the dimensionality d of the Hilbert space, the most complete measurements being those that resolve the Hilbert space into d 1-dimensional subspaces.

Let $M_k$ represent the projection operator onto the k'th subspace of measurement M, so that the identity operator on H can be represented as a sum of projections: $I = M_1 + M_2 + \ldots$. When a system in state $\psi$ is subjected to measurement M, its behavior is in general probabilistic: outcome k occurs with a probability equal to $|M_k \psi|^2$, the square of the length of the state vector's projection into subspace $M_k$. After the measurement, the system is left in a new state $M_k\psi/|M_k\psi|$, which is the normalized unit vector in the direction of the old state vector's projection into subspace $M_k$. The measurement thus has a deterministic outcome, and leaves the state vector unmodified, only in the exceptional case that the initial state vector happens to lie entirely in one of the orthogonal subspaces characterizing the measurement.

The Hilbert space for a single polarized photon is 2-dimensional; thus the state of a photon may be completely described as a linear combination of, for example, the two unit vectors $r_1 = (1,0)$ and $r_2 = (0,1)$, representing respectively horizontal and vertical polarization. In particular, a photon polarized at angle $\alpha$ to the horizontal is described by the state vector $(\cos\alpha, \sin\alpha)$. When subjected to a measurement of vertical-vs.-horizontal polarization, such a photon in effect chooses to become horizontal with probability $\cos^2\alpha$ and vertical with probability $\sin^2\alpha$. The two orthogonal vectors $r_1$ and $r_2$ thus a exemplify the resolution of a 2-dimensional Hilbert space into 2 orthogonal 1-dimensional subspaces; henceforth $r_1$ and $r_2$ will be said to comprise the 'rectilinear' basis for the Hilbert space.

An alternative basis for the same Hilbert space is provided by the two 'diagonal' basis vectors $d_1 = (0.707, 0.707)$, representing a 45-degree photon, and $d_2 = (0.707, -0.707)$, representing a 135-degree photon. Two bases (e.g. rectilinear and diagonal) are said to be 'conjugate' [W], if each vector of one basis has equal-length projections onto all vectors of the other basis: this means that a system prepared in a specific state of one basis will behave entirely randomly, and lose all its stored information, when subjected to a measurement corresponding to the other basis. Owing to the complex nature of its coefficients, the two-dimensional Hilbert space also admits a third basis conjugate to both the rectilinear and diagonal bases, comprising the two so-called 'circular' polarizations $c_1 = (0.707, 0.707i)$ and $c_2 = (0.707i, 0.707)$; but the rectilinear and diagonal bases are all that will be needed for the cryptographic applications in this paper.

The Hilbert space for a compound system is constructed by taking the tensor product of the Hilbert spaces of its components; thus the state of a pair of photons is characterized by a unit vector in the 4-dimensional Hilbert space spanned by the orthogonal basis vectors $r_1 r_1$, $r_1 r_2$, $r_2 r_1$, and $r_2 r_2$. This formalism entails that the state of a compound system is not generally expressible as the cartesian product of the states of its parts: e.g. the Einstein-Podolsky-Rosen state of two photons, $0.7071(r_1 r_2 - r_2 r_1)$, to be discussed later, is not equivalent to any product of one-photon states.

III. Quantum Public Key Distribution

In traditional public-key cryptography, trapdoor functions are used to conceal the meaning of messages between two users from a passive eavesdropper, despite the lack of any initial shared secret information between the two users. In quantum public key distribution, the quantum channel is not used directly to send meaningful messages, but is rather used to transmit a supply of random bits between two users who share no secret information initially, in such a way that the users, by subsequent consultation over an ordinary non-quantum channel subject to passive eavesdropping, can tell with high probability whether the original quantum transmission has been disturbed in transit, as it would be by an eavesdropper (it is the quantum channel's peculiar virtue to compel eavesdropping to be active). If the transmission has not been disturbed, they agree to use these shared secret bits in the well-known way as a one-time pad to conceal the meaning of subsequent meaningful communications, or for other cryptographic applications (e.g. authentication tags) requiring shared secret random information. If transmission has been disturbed, they discard it and try again, deferring any meaningful communications until they have succeeded in transmitting enough random bits through the quantum channel to serve as a one-time pad.

In more detail one user ('Alice') chooses a random bit string and a random sequence of polarization bases (rectilinear or diagonal). She then sends the other user (Bob) a train of photons, each representing one bit of the string in the basis chosen for that bit position, a horizontal or 45-degree photon standing for a binary zero and a vertical or 135-degree photon standing for a binary 1. As Bob receives the photons he decides, randomly for each photon and independently of Alice, whether to measure the photon's rectilinear polarization or its diagonal polarization, and interprets the result of the measurement as a binary zero or one. As explained in the previous section a random answer is produced and all information lost when one attempts to measure the rectilinear polarization of a diagonal photon, or vice versa. Thus Bob obtains meaningful data from only half the photons he detects--those for which he guessed the correct polarization basis. Bob's information is further degraded by the fact that, realistically, some of the photons would be lost in transit or would fail to be counted by Bob's imperfectly-efficient detectors.

Subsequent steps of the protocol take place over an ordinary public communications channel, assumed to be susceptible to eavesdropping but not to the injection or alteration of messages. Bob and Alice first determine, by public exchange of messages, which photons were successfully received and of these which were received with the correct basis. If the quantum transmission has been undisturbed, Alice and Bob should agree on the bits encoded by these photons, even this data has never been discussed over the public channel. Each of these photons, in other words, presumably carries one bit of random information (e.g. whether a rectilinear photon was vertical or horizontal) known to Alice and Bob but to no one else.

Because of the random mix of rectilinear and diagonal photons in the quantum transmission, any eavesdropping carries the risk of altering the transmission in such a way as to produce disagreement between Bob and Alice on some of the bits on which they think they should agree. Specifically, it can be shown that no measurement on a photon in transit, by an eavesdropper who is informed of the photon's original basis only after he has performed his measurement, can yield more than 1/2 expected bits of information about the key bit encoded by that photon; and that any such measurement yielding b bits of expected information (b ≤ 1/2) must induce a disagreement with probability at least b/2 if the measured photon, or an attempted forgery of it, is later re-measured in its original basis. (This optimum tradeoff occurs, for example, when the eavesdropper measures and retransmits all intercepted photons in the rectilinear basis, thereby learning the correct polarizations of half the photons and inducing disagreements in 1/4 of those that are later re-measured in the original basis.)

Alice and Bob can therefore test for eavesdropping by publicly comparing some of the bits on which they think they should agree, though of course this sacrifices the secrecy of these bits. The bit positions used in this comparison should be a random subset (say one third) of the correctly received bits, so that eavesdropping on more than a few photons is unlikely to escape detection. If all the comparisons agree, Alice and Bob can conclude that the quantum transmission has been free of significant eavesdropping; and those of the remaining bits that were sent and received with the same basis also agree, and can safely be used as a one time pad for subsequent secure communications over the public channel. When this one-time pad is used up, the protocol is repeated to send a new body of random information over the quantum channel.

The following example illustrates the above protocol.

| QUANTUM TRANSMISSION | | | | | | | | | | | | | | |
|---|---|---|---|---|---|---|---|---|---|---|---|---|---|---|
| Alice's random bits........................ ∅ | 1 | 1 | 0 | 1 | 1 | 0 | 0 | 1 | 0 | 1 | 1 | 0 | 0 | 1 |
| Random sending bases....................... D | R | D | R | R | R | R | R | D | D | R | D | D | D | R |
| Photons Alice sends........................ ↗ | ↕ | ↖ | ↔ | ↕ | ↕ | ↔ | ↔ | ↖ | ↗ | ↕ | ↖ | ↗ | ↗ | ↕ |
| Random receiving bases..................... R | D | D | R | R | D | D | R | D | R | D | D | D | D | R |
| Bits as received by Bob.................... 1 | | 1 | | 1 | 0 | 0 | 0 | | 1 | 1 | 1 | | 0 | 1 |
| PUBLIC DISCUSSION | | | | | | | | | | | | | | |
| Bob reports bases of received bits......... R | D | | R | | D | D | R | | R | D | D | | D | R |
| Alice says which bases were correct........ | | | OK | | OK | | OK | | | OK | | | OK | OK |
| Presumably shared information (if no eavesdrop)... | | | 1 | | 1 | | 0 | | | 1 | | | 0 | 1 |
| Bob reveals some key bits at random........ | | | | | 1 | | | | | | | | 0 | |
| Alice confirms them........................ | | | | | OK | | | | | | | | OK | |
| OUTCOME | | | | | | | | | | | | | | |
| Remaining shared secret bits............... | | | 1 | | | | 0 | | | 1 | | | | 1 |

The need for the public (non-quantum) channel in this scheme to be immune to *active* eavesdropping can be relaxed if the Alice and Bob have agreed beforehand on a small secret key, which they use to create Wegman-Carter authentication tags [WC] for their messages over the public channel. In more detail the Wegman-Carter multiple-message authentication scheme uses a small random key to produce a message-dependent 'tag' (rather like a check sum) for an arbitrary large message, in such a way that an eavesdropper ignorant of the key has only a small probability of being able to generate any other valid message-tag pairs. The tag thus provides evidence that the message is legitimate, and was not generated or altered by someone ignorant of the key. (Key bits are gradually used up in the Wegman-Carter scheme, and cannot be reused without compromising the system's provable security; however, in the present application, these key bits can be replaced by fresh random bits successfully transmitted through the quantum channel.) The eavesdropper can still prevent communication by suppressing messages in the public channel, as of course he can by suppressing or excessively perturbing the photons sent through the quantum channel. However, in either case, Alice and Bob will conclude with high probability that their secret communications are being suppressed, and will not be fooled into thinking their communications are secure when in fact they're not.

IV. Quantum Coin Tossing

'Coin Flipping by Telephone' was first discussed by Blum [B1]. The problem is for two distrustful parties, communicating at a distance without the help of a third party, to come to agree on a winner and a loser in such a way that each party has exactly 50 per cent chance of winning. Any attempt by either party to bias the outcome should be detected by the other party as cheating. Previous protocols for this problem are based on unproved assumptions in computational complexity theory, which makes them vulnerable to a breakthrough in algorithm design.

By contrast, we present here a scheme involving classical and quantum messages which is secure against traditional kinds of cheating, even by an opponent with unlimited computing power. Ironically, it can be subverted by a still subtler quantum phenomenon, the so-called Einstein-Podolsky-Rosen effect. This threat is merely theoretical, because it requires perfect efficiency of storage and detection of photons, which though not impossible in principle is far beyond the capabilities of current technology. The honestly-followed protocol, on the other hand, could be realized with current technology.

1. Alice chooses randomly one basis (say rectilinear) and a sequence of random bits (one thousand should be sufficient). She then encodes her bits as a sequence of photons in this same basis, using the same coding scheme as before. She sends the resulting train of polarized photons to Bob.

2. Bob chooses, independently and randomly for each photon, a sequence of reading bases. He reads the photons accordingly, recording the results in two tables, one of rectilinearly received photons and one of diagonally received photons. Because of losses in his detectors and of the transmission channel, some of the photons may not be received at all, resulting in holes in his tables. At this time, Bob makes his guess as to which basis Alice used, and announces it to Alice. He wins if he guessed correctly, loses otherwise.

3. Alice reports to Bob whether he won, by telling him which basis she had actually used. She certifies this information by sending Bob, over a classical channel, her entire original bit sequence used in step 1.

4. Bob verifies that no cheating has occurred by comparing Alice's sequence with both his tables. There should be perfect agreement with the table corresponding to Alice's basis and no correlation with the other table. In our example, Bob can be confident that Alice's original basis was indeed rectilinear as claimed.

Illustrating the protocol by a specific example,

| | | | | | | | | | | | | | | |
|---|---|---|---|---|---|---|---|---|---|---|---|---|---|---|
| Alice's bit string..................................... | 1 | 0 | 1 | 0 | 0 | 1 | 1 | 1 | 0 | 1 | 0 | 1 | 1 | 0 | 0 |
| Alice's random basis................................ | | | | | | Rectilinear | | | | | | | | | |
| Photons Alice sends.................................. | ↕ | ↔ | ↕ | ↔ | ↔ | ↕ | ↕ | ↕ | ↔ | ↕ | ↔ | ↕ | ↕ | ↔ | ↔ |
| Bob's random bases.................................. | R | D | D | D | R | R | D | R | R | D | R | R | D | D | R |
| Bob's rectilinear table............................. | 1 | | | | 1 | 1 | | 1 | 0 | | 1 | 1 | | | 0 |
| Bob's diagonal table................................ | | 0 | 1 | | | | 1 | | | 0 | | | 1 | 0 | |
| Bob's guess............................................. | | | | | | 'Rectilinear' | | | | | | | | | |
| Alice's reply.......................................... | | | | | | 'You win' | | | | | | | | | |
| Alice sends her original bit string to certify.... | 1 | 0 | 1 | 0 | 0 | 1 | 1 | 1 | 0 | 1 | 0 | 1 | 1 | 0 | 0' |
| Bob's rectilinear table............................. | 1 | | | | 1 | 1 | | 1 | 0 | | 1 | 1 | | | 0 |
| Bob's diagonal table................................ | | 0 | 1 | | | | 1 | | | 0 | | | 1 | 0 | |

In order to cheat, Bob would need to guess Alice's basis with probability greater than 1/2. This amounts to distingushing a train of photons randomly polarized in one basis from a train randomly polarized in another basis. However, it can be shown that any measuring apparatus capable of making this distinction can also be used, in conjunction with the Einstein-Podolsky-Rosen effect described below to transmit useful information faster than the speed of light, in violation of well-established physical laws.

Alice could attempt cheating either at step 1 or step 2. Let us first assume that she follows step 1 honestly and finds herself losing at the end of step 2, because Bob made he correct guess, here rectiliniear. In order to pretend she has won, she would need to convince Bob that her photons were diagonally polarized, which she can only do by producing a sequence of bits in perfect agreement with Bob's diagonal table. This she cannot do reliably because this table is the result of probabilistic behavior of the photons after the left her hands. Suppose she goes ahead anyway and sends Bob a new 'original' sequence, different from the one that she used in step 1, in hopes that it will by luck agree perfectly with Bob's diaognal table. This attempt to cheat requires Alice to be not only lucky but daring, because in the vast majority of cases, the gamble would fail and would be detected as cheating. By contrast, in traditional coin-tossing schemes, analogous attempts to seize a lucky victory from the jaws of defeat, though unlikely to succeeed, are unaccompanied by any danger of detection.

It is easy to see that things are even worse for Alice if she attempts to cheat in step 1, by sending a mixture of rectilinear and diagonal photons, or photons which are polarized neither rectilinearly or diagonally. In this case she will not be able to agree with either of Bob's tables in step 3, since both tables will record the results of probabilistic behavior not under her control.

In order to say how Alice *can* cheat using quantum mechanics it is necessary to describe the Einstein-Podolsky-Rosen (EPR) effect [Bo,AGR], often called a paradox because it contradicts the common-sense notion that for two individually random events happening at distance from one another to be correlated, some physical influence must have propagated from the earlier event to the later, or else from some common random cause to both events.

The EPR effect occurs when certain types of atom or molecule decay with the emission of two photons, and consists of the fact that the two photons are always found to have opposite polarization, regardless of the basis used to observe them, provided both are observed in the same basis. For example, if both photons are measured rectilinearly, it will always be found that one is horizontal and the other vertical, though which is horizontal will vary randomly from one decay to the next. If both photons are measured diagonally, one will always be 135-degree and the other 45-degree. A moment's reflection will show that this behavior cannot be explained by assuming the decay produces a distribution over $\alpha$ of oppositely polarized ($\alpha$ and $\alpha+90$) photons, since, in that case, if such a pair of photons were measured in an intermediate basis (say $\alpha+45$), both would behave probabilistically so as to sometimes come out with the same polarization.

Probably the simplest, but paradoxical-sounding, verbal explanation of the EPR effect is to say that the two photons are produced in an initial state of undefined polarization; and when one of them is measured, the measuring apparatus forces it to choose a polarization (choosing randomly and equiprobably between the two characteristic directions offered by the apparatus) while simultaneously forcing the other unmeasured photon, no matter how far away, to choose the opposite polarization. This implausible-sounding explanation is supported by formal quantum mechanics, which represents the state of a pair of photons as a vector in a 4-dimensional Hilbert space obtained by taking the tensor product of two 2-dimensional Hilbert spaces. The EPR state produced by the decay is described by the vector $0.7071(r_1 r_2 - r_2 r_1)$, and the EPR effect is explained by the fact that this vector has anticorrelated projections into the 2-dimensional Hilbert spaces of the two photons no matter what basis is used to express the tensor product (e.g. the same state vector is demonstrably equal to $0.7071(d_1 d_2 - d_2 d_1)$, and to $0.7071(c_1 c_2 - c_2 c_1)$).

In order to cheat, Alice produces a number of EPR photon-pairs instead of individual random photons in step 1. In each case she sends Bob one member of the pair and stores the other herself, perhaps between perfectly reflecting mirrors. When Bob makes his guess (e.g. rectilinear) she then measures all her stored photons in the opposite (diagonal) basis, thereby obtaining results perfectly correlated with his diagonal table but uncorrelated with his rectilinear table. She then announces these results, pretending them to be the random bits she was supposed to have encoded in the photons in step 1, and thereby forces a win from which Bob cannot escape even by delaying his measurements until after his guess. This cheat requires that Alice be able to store the twin photons for a considerable time and then measure them with high detection efficiency, and thus would be possible only in principle, not in practice. Any photons lost by Alice during storage or measurement would result in holes in her pretended bit sequence, which she would have to fill by guessing, and these guesses would risk detection by Bob if they failed to agree with his tables.

REFERENCES

[AGR] A. Aspect, P. Grangier, and G. Roger, 'Experimental Realization of the Einstein-Podolsky-Rosen-Bohm Gedankenexperiment: a New Violation of Bell's Inequalities', *Phys.Rev.Lett.* 49, 91-94 (1982).

[BBBW] C.H.Bennett, G. Brassard, S.Breidbart, and S. Wiesner, 'Quantum Cryptography, or Unforgeable Subway Tokens', to appear in *Advances in Cryptography: Proceedings of CRYPTO82*, Plenum Press.

[Bl] Manuel Blum, 'Coin Flipping by Telephone-- a Protocol for Solving Impossible Problems', *SIGACT News* 15:1, 23-27 (1983).

[Bo] David Bohm, *Quantum Theory* (Prentice-Hall, Englewood Cliffs, NJ 1951), pp. 614-619.

[WC] M. Wegman and L.Carter, 'New Hash Functions and Their Use in Authentication and Set Equality,' *J.Comp.Sys.Sci.* 22, 265-279, (1981).

[W] Stephen Wiesner, 'Conjugate Coding', (manuscript ca 1970); subsequently published in *SIGACT News* 15:1, 78-88 (1983).

[WZ] W.K. Wootters and W.H. Zurek, 'A Single Quantum Cannot be Cloned', *Nature* 299, 802-803 (1982).